\documentclass{iopart}
\usepackage{epsfig}
\begin{document}

\begin{flushright}
TTP99-04
\end{flushright}
\title{Spin physics in deep inelastic scattering: Summary}
\author{T Gehrmann\dag, T Sloan\ddag}
\address{\dag\ Institut f\"ur Theoretische Teilchenphysik, 
Universit\"at Karlsruhe, D-76128 Karlsruhe, Germany}
\address{\ddag\ University of Lancaster,
Lancaster   LA1 4YB, UK,
E-mail: t.sloan@lancaster.ac.uk}
\vspace{0.7cm}

\begin{abstract}
The problem of our understanding of the spin  structure of the nucleon has
been with us since the publication of the EMC measurements of the polarised
structure function of the proton in 1987.  In this talk a review of the 
results presented in Working Group 6 at this workshop is given. 
\end{abstract}

\section{Brief History}

In the simple quark model the  spin of the proton is carried by
its three valence quarks so that $\Delta\Sigma= \Delta u + \Delta d = 1$.
Here $\Delta q$ = $\int^1_0 dx\, \left(q_{\uparrow}(x)-q_{\downarrow}(x)
\right)$   where
$q_{\uparrow(\downarrow)}(x)$ 
are distributions for quarks with spin aligned (anti-aligned) to the
proton spin and $q$ = $u, d$, etc.\ indicates the quark flavours.  
The simple quark model has however proven to be inadequate long before
precise measurements of the proton spin structure became available,
since
it predicts that the ratio of the axial vector to vector coupling
constants in neutron $\beta$ decay is $g_A = 5/3$ compared to the
measured value of 1.26. 

The parton model ascribes part of the proton spin to 
sea-quarks and gluons. All partons in the proton can moreover possess 
orbital angular momentum, which also contributes to the proton spin.
Within this model, the proton spin can no longer be identified with  the
sum of the quark spins only, and 
$\Delta \Sigma$ can therefore not be predicted without making additional
assumptions. The best-known theoretical prediction of $\Delta \Sigma$ is
due to Ellis and Jaffe~\cite{EJ}. Using SU(3)-flavour symmetry  
with the additional assumption of vanishing of the contribution from 
strange quarks to the proton spin, they obtain $\Delta \Sigma \approx
0.58$. 

Deep inelastic scattering (DIS) with polarised
charged leptons on polarised targets allows the quark distributions
$q_{\uparrow(\downarrow)}$  to be investigated.  These are extracted from 
the structure function $g_1(x,Q^2)$ measured in polarised DIS 
using the parton model relation
$g_1(x,Q^2) =\frac{1}{2}\sum_q
e_q^2\,\left(q_{\uparrow}(x)-q_{\downarrow}(x)
\right)$.
In the early 1980s the SLAC experiments E80 and E130~\cite{SLAC}
reported the first measurements of polarised DIS for $x > 0.1$.
In 1988, the EMC reported measurements~\cite{EMC} over a range
down to $x = 0.015$. For $x > 0.1$
all the data (extrapolated to $x=0$ for the determination of $\Delta \Sigma$)
seemed to confirm the expectations of Ellis and Jaffe.
However, as $x$ decreased the EMC data fell progressively
below 
the expectations of the quark parton model and 
yielded a very small $\Delta \Sigma$, which 
was even consistent with zero at that time. 
The value of $\Delta \Sigma$ has increased since then due to the 
refinement of our knowledge of F and D, the SU(3) couplings measured in 
hyperon beta decay (see~\cite{Ratcliffe} for the latest analysis).  
However, there is still a significant difference of the measurement 
from the value expected from the Ellis-Jaffe sum rule.  
The significance of this disagreement implies that only a small 
fraction of the spin of the proton is carried by quark spins. 

This surprising result created great theoretical interest. Where was the
spin of the nucleon ?  Could it be in the gluons ($\Delta g$)
as suggested in~\cite{EandT,AandR} or could it
be in orbital angular momentum ($L_q$, $L_g$)~\cite{[5.]}. By angular momentum
conservation the total spin of the nucleon of 1/2 must be equal to 1/2
$\Delta \Sigma +\Delta g + L_q + L_g$.  It was also
suggested that the problem did not exist and part of $\Delta \Sigma$
was missed in the unmeasured region at very small $x$~\cite{CandR}.
All this interest motivated a new experimental programme to investigate the
phenomenon further and this programme is now coming to fruition. 

On the theoretical side, much confusion was caused by the scheme
dependence of $\Delta \Sigma$ in higher orders of perturbative QCD. 
This problem could only be resolved three years ago with the calculation 
of the two-loop polarized splitting functions,~\cite{twoloop}, now
allowing to define consistent transformations between different
factorization schemes~\cite{ridolfi}. 
In the recent past, the next-to-leading order
QCD corrections for the majority
 of the experimentally relevant polarized observables have been
 calculated, see for example~\cite{BT} for a review. 

\section{Recent Experimental Results}

The SMC has presented data over the widest range of $x$ on the polarised
structure functions~\cite{BT,EWA}.  The collaboration has greatly improved
the precision of the data at low $x$ by demanding an observed hadron in 
each event.  This rejects radiative  
and other events with low depolarisation factors.  The remaining events 
are then undiluted by data of poor significance for the asymmetry 
determination allowing the asymmetry to be measured more precisely.
Furthermore, a much lower $Q^2$ trigger has been implemented which allows
asymmetries to be measured in the range $10^{-4} < x < 10^{-3}$.  The data
from this trigger serve to investigate the Regge region to search for a
possible divergence at low $x$ such as 
proposed in~\cite{CandR}.  Fig.~1 shows the SMC data~\cite{Kiryluk}  with
the behaviour of $g_1 = 0.17/x\,\ln^2 x$ (solid curve) proposed
by~\cite{CandR}.  Such behaviour is now excluded by the data.  
 However, the 
less extreme behaviours $g_1 =-0.14\,\ln x$ (dashed curve) and 
$g_1 = -0.085(2+\ln x)$ (dotted curve) which were also proposed 
in~\cite{CandR} cannot be excluded.  All the curves 
were calculated assuming a value of $R=\sigma_L/\sigma_T=0$. Hence they 
represent lower limits since the curves scale as $1+R$.  The first of these 
behaviours would make a sizable difference to the determination of 
$\Delta\Sigma$ so its exclusion removes a significant uncertainty.  

Direct comparison of these data with the double 
logarithmic small-$x$ resummations of~\cite{Ryskinetal} is difficult due 
to the low $Q^2$ values involved. A model for extending these
resummations into the low $Q^2$ region is discussed in detail in
these proceedings~\cite{badelek}, it is in good agreement with the data. 
\begin{figure}[ht]
\begin{center}
\epsfig{file=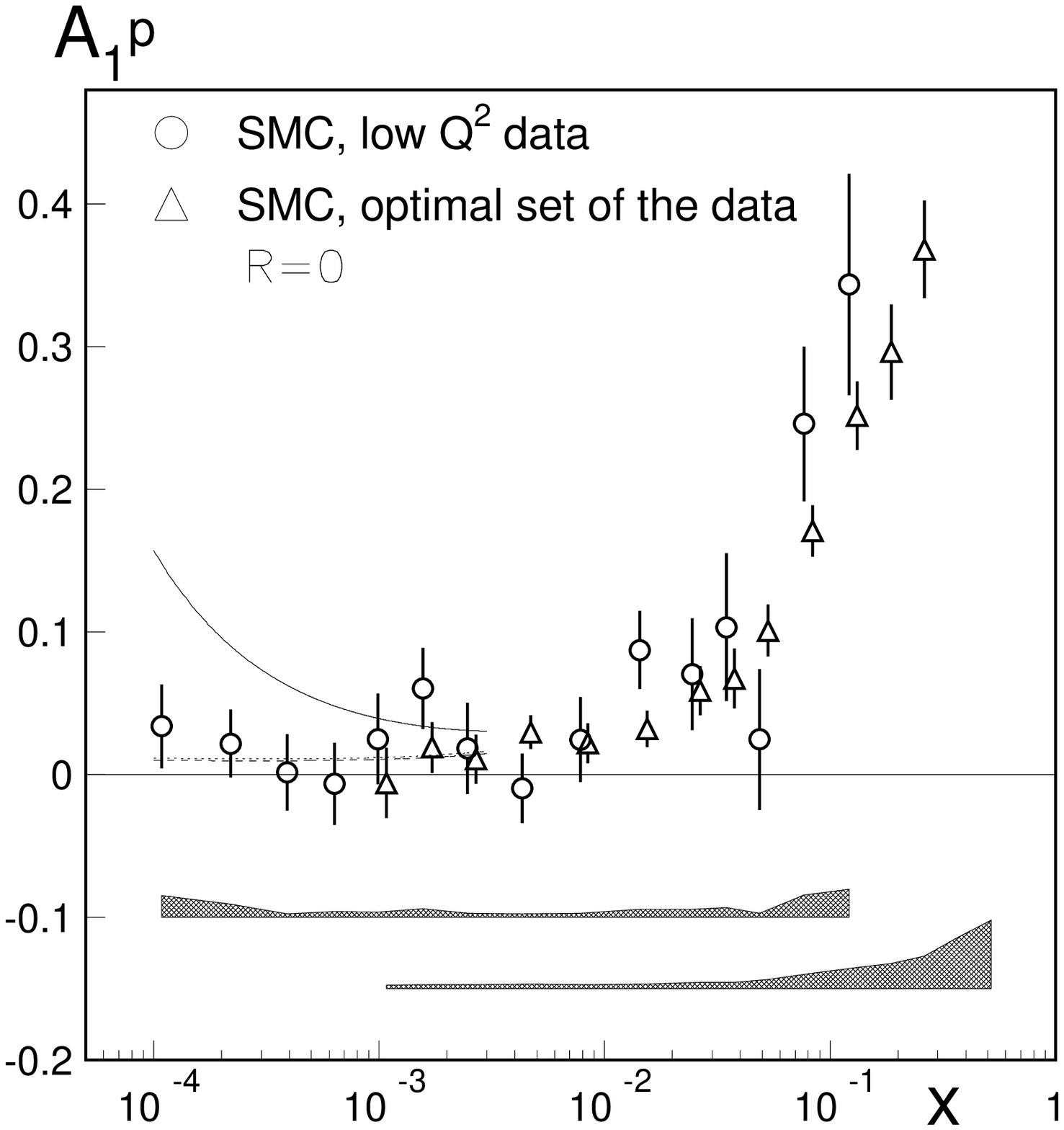,width=9cm}
\end{center}
{\footnotesize
Figure 1. The values of $A_1^p$ as a function of $x$ measured by 
SMC~\cite{Kiryluk} (preliminary).  
The smooth curves show the expected behaviour of $A_1^p$ 
as $x  \rightarrow 0$ proposed in ~\cite{CandR}.  The solid curve 
shows the behaviour for $g_1^p \sim 1/x\,\ln^2 x$, the dashed curve 
for $g_1 \sim \ln x$ and the dotted curve $g_1 \sim (2+\ln x)$.  }
\end{figure}

The SMC group have made NLO QCD fits to the world data 
in an attempt to determine $\Delta g$, \cite{BT,EWA,LEADER}.
The theoretical error on this quantity
can be estimated by varying renormalization 
and factorization scales in the fits~\cite{ridolfi}. 
These variations generate terms which 
are compensated only in the NNLO order expressions,
such that the resulting error can be taken as a measure of the
importance of higher orders in the perturbative series.
In~\cite{EWA} this error  
is assessed by varying the scales between 
the limits of $Q^2/2$ and $2 Q^2$, 
resulting in a rather large variation of $\Delta g$.
Given that most of the data included in the fit are at moderate 
$Q^2 \sim 1\ldots 10$~GeV$^2$, one would indeed assume that perturbative 
corrections beyond NLO (as well as target mass corrections~\cite{avto})
could be sizable. It should however be pointed
out that theoretical error and statistical error on 
$\Delta g$ are of a similar magnitude, such that improvement on the
theoretical side only would not be sufficient for a better determination 
of $\Delta g$. This clearly illustrates the necessity for a direct 
measurement of this quantity.

Interesting recent results have also been reported to this workshop 
from HERMES~\cite{BT} 
in which the semi-inclusive distributions 
of charged hadrons have been used to deduce the 
parton distributions for 
individual quark flavours to the spin of the proton. These data add to earlier
SMC measurements~\cite{SMChad}.  
Upgrades to the HERMES detector will soon  
allow separation of different hadron species, which might yield the
first flavour decomposition of the light quark sea~\cite{BT}.

\section{Theoretical Progress}

A consistent extraction of parton distributions at next-to-leading order 
requires knowledge of both NLO splitting functions and subprocess cross
sections for all experimental observables included in a global fit. Up
to now, these fits were restricted to structure function measurements
only. However, the range of polarized observables will soon be extended 
with a variety of new reactions to be measured at COMPASS and 
RHIC~\cite{futspin}. 
For many of these, subprocess cross sections are now available at NLO.  

Most recently, NLO corrections to the photoproduction 
of heavy quarks have been calculated~\cite{strat}.
An important outcome of this calculation is the relative smallness of 
light quark induced contributions in photoproduction of charm. The
considerably improved dependence on factorization and renormalization
scale at next-to-leading order indicates moreover the perturbative
stability of this observable, which can therefore be used for a reliable 
determination of $\Delta g$ once data become available. 

First progress towards the calculation of the polarized splitting
functions at NNLO has been reported by Gracey~\cite{gracey}. Using the 
$1/N_f$ expansion, several terms of the polarized splitting functions
could be determined to all orders. These results could serve as a
consistency check once full results for the splitting functions become
available. 
Another important test of higher  order corrections are the
relations between polarized and unpolarized results: these are discussed 
in~\cite{ravi}.

Presently,
the contribution of partonic angular momentum to the proton spin 
is not at all determined. Using the recently derived renormalization group
equations for the angular momentum distributions~\cite{haegler}, it is
now feasible to model these distributions. 

The behaviour of the polarized proton structure at small $x$ is 
expected to be governed by leading double logarithmic terms of the form 
$\alpha_s^n \ln^{2n} x$, which are absent in the 
unpolarized singlet structure functions. A resummation of these terms
has been performed in~\cite{Ryskinetal}, and their 
impact has been the topic of extensive discussions during
the workshop. It is commonly agreed that the effect of 
the small-$x$ resummation can not be 
tested on the current small-$x$ data from SMC~\cite{Kiryluk},
 which correspond to  only very
low photon virtualities. A model for $g_1$ at low $Q^2$ and small $x$
incorporating  resummation was proposed by Badelek and Kwieci\'{n}ski and 
yields a decent description of the experimental data~\cite{badelek}.
Further observables studied in the same double logarithmic framework
 are $g_2$ at small
$x$~\cite{g2} and the diffractive content of $g_1$~\cite{g1d}, which are 
however inaccessible at present experiments. Like in the unpolarized
case, decent probes of phenomena at small $x$ would only be possible
with a polarized electron-proton collider, such as the currently discussed
polarized HERA option.

In inclusive DIS, the measurement of the polarized gluon distribution
is indirect. Various more direct measurements of  $\Delta g$ have 
proposed such as the observation of charm production at COMPASS
and in HERMES as well as di-jet production using polarised 
protons in RHIC and in HERA~\cite{deroeck}.  A problem 
which was discussed extensively at the workshop was the associated production 
of charmed baryons and mesons.  First Monte Carlo studies 
based on the LUND string model~\cite{gerdt} indicate that 
such backgrounds may not be serious at COMPASS energies but could become 
substantial at HERMES energies.  As a result of the workshop, more
involved theoretical studies have been carried out. Modeling associated 
production as interchange of constituent quarks, Ryskin and Leader
confirm~\cite{ryskin} that an open charm production measurement at HERMES 
will suffer from a large contamination due to associated production,
such that it will not yield conclusive information on $\Delta g$. 

Another potential probe of the polarized gluon distribution is the
photoproduction of $J/\psi$ mesons. The inelastic production is induced
by boson-gluon fusion, and thus directly proportional to the gluon
distribution. Under realistic experimental conditions, it is however
very hard to separate inelastic from elastic production. A decent 
theoretical description of unpolarized inelastic $J/\psi$ production is
given in the perturbative two gluon exchange model of~\cite{ryskinjpsi}.
 This model has now 
been applied by Mankiewicz and V\"anttinen~\cite{mv} to compute
production asymmetries in elastic $J/\psi$ production. Contrary to
earlier claims in the literature, it could be proven that the elastic
$J/\psi$ production cross section is insensitive to the spin states of
probe and target, a small spin dependence is induced only from 
relativistic corrections. As a consequence, elastic $J/\psi$ production 
can not be used to probe the polarized gluon distribution, as initially
hoped. 

\section{Conclusions}

Ten years after the release of the EMC measurement of the small
contribution of quark spins to the proton spin, an extensive amount of 
spin structure function measurements is available. Confirming the 
initial EMC observation, these measurements have contributed much
information on the polarized quark distributions in the proton. 
Our
picture of the proton spin structure is however far form being complete: 
current data yield only loose constraints on the polarized gluon
distribution, and no information is available yet on angular momentum 
contributions to the proton spin. 

The theoretical understanding of the spin structure of the nucleon has
vastly improved, with the large majority of accessible observables now
being calculated to NLO. Theoretical efforts are now extending in
various directions: understanding of spin effects at small $x$,
computation of NNLO corrections and investigation of angular momentum
distributions are examples of currently ongoing research work.  

Making further experimental
progress towards a determination of $\Delta g$ seems 
to be harder than originally anticipated. Concerning the prospects of 
extracting $\Delta g$ from charm production at HERMES energies, the
working group has concluded that neither elastic $J/\psi$ production
(vanishing asymmetry at partonic level)  
nor open charm production (large background from associated production)
are reliable channels. A measurement from open charm production
at COMPASS energies looks far more promising due to the much reduced 
background. With the recently calculated NLO corrections, the
theoretical uncertainties of this observable appear also to be under
control. 

In addition to COMPASS, other future experiments promise to yield new valuable
information on the nucleon's spin structure. A whole range of new
observables will become accessible at the RHIC 
polarized proton-proton collider, which is currently constructed
at BNL. The option of polarizing the HERA proton beam, 
which is under extensive study for the moment, would largely extend the
kinematical region covered by present fixed target experiments
and allow 
to study a variety of new channels probing the spin structure of both 
photon and proton.

\section*{Acknowledgments}
We thank the organisers for creating such an interesting, stimulating
and enjoyable workshop.  We also thank all the participants in Working Group
6 for their assistance in preparing this talk.

\end{document}